\documentstyle[twocolumn,pre,aps,epsf,floats,psfig]{revtex}

\newcommand{\nn}{\nonumber}

\begin{document}
 
\twocolumn[\hsize\textwidth\columnwidth\hsize\csname @twocolumnfalse\endcsname 
 
\date{\today}
\title{Triangular anisotropies in Driven Diffusive Systems: 
reconciliation of Up and Down}
\author{A. D. Rutenberg$^{1}$ and C. Yeung$^{2}$}
\address{
$^{1}$ Centre for the Physics of Materials, Physics Department,
McGill University, Montr\'{e}al QC, Canada H3A 2T8 \\
$^{2}$ School of Science, Pennsylvania State University at Erie, The
Behrend College, Erie PA, 16563 USA} 
\maketitle

\widetext

\begin{abstract}
Deterministic coarse-grained descriptions of driven diffusive systems (DDS) 
have been hampered by apparent inconsistencies with kinetic
Ising models of DDS.
In the evolution towards the driven steady-state,
``triangular'' anisotropies in the two systems point in 
opposite directions with respect to the drive field.   We show that this
is non-universal behavior in the sense that the triangular anisotropy
``flips'' with local modifications of the Ising interactions.
The sign and magnitude of the triangular anisotropy also vary with temperature. 
We have also flipped the anisotropy of coarse-grained models, though not yet
at the latest stages of evolution.  Our results illustrate the 
comparison of deterministic coarse-grained and stochastic Ising DDS studies to 
identify universal phenomena in driven systems. 
Coarse-grained systems are particularly attractive in terms
of analysis and computational efficiency.

\end{abstract}
\pacs{64.60.My,05.60.+w,66.30.Hs,05.50.+q,64.60.Cn}

\narrowtext ]                   


Driven steady-state systems are common in many fields
of physics including device physics and materials processing, and are also
found in many biological processes.  The simplest
example of a driven system is one in which a uniform external drive is
applied.   For closed boundary conditions, a final equilibrium state is reached
in which the external drive is balanced by other forces --- as in 
a closed system in a gravitational field.  For open
boundaries, equilibrium will never be reached and a non-equilibrium
steady state will continue ``indefinitely''. Electrical circuits provide
prosaic examples of this. In both open and closed systems, the introduction
of local interactions provide a rich phenomenology that is only slowly being 
explored.  

To  model a driven system we must describe both the energetics and the 
dynamics.  One of the simplest
such model is a lattice driven diffusive system (DDS).  This is simply an
interacting lattice gas with biased motion in
the direction of the uniform external field.  
Early work on this model concentrated on steady-state properties
near the critical point, which survives from the underlying zero-field Ising
model.  More recently there has been increasing emphasis on the
approach to the steady state. This dynamical regime is independent
of the open or closed boundary conditions, and hence is common to both.
See the book by Schmittmann and Zia \cite{Schmittmann95} for an
introduction to the literature.  

One of the most important questions we can ask about any model is whether the
behavior that it displays is universal. To whit: is the observed 
behavior seen in a broad class of systems, or is it specific to that precise
model? For example, we can ask whether a lattice DDS  and related
coarse-grained models display the same universal behavior.  This question
has proven remarkably controversial in DDS, and this paper aims
towards a reconciliation between lattice and coarse-grained
approaches.  

\begin{figure}

\centerline{ \psfig{figure=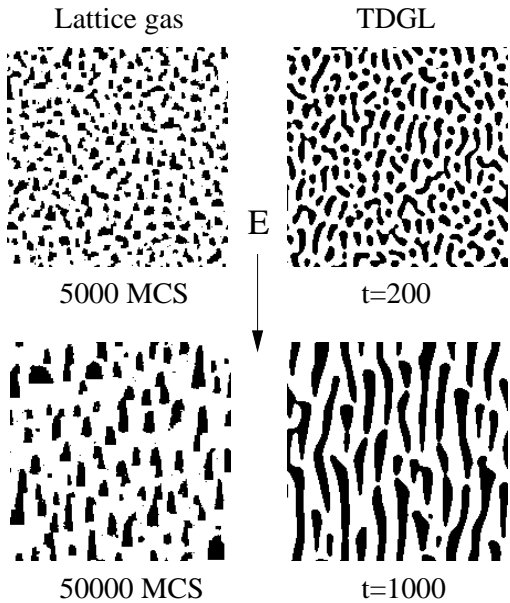,width=2.625in} } 

\ \\
\caption{
\label{FIG:offcrit}
Evolution of domains for an off-critical quench in
a lattice gas  with nearest neighbor hops (left) and for a TDGL 
coarse-grained model with isotropic mobility (right).  The field ${\bf E}$
biases the motion of the particles (dark) downwards.
For the lattice gas the tips of the triangular domains point ``up'', 
opposite the field, for the coarse-grained model the domains point ``down''.}

\end{figure}

Near the critical point separating 
the low-temperature ordered and high-temperature disordered DDS phases, 
coarse-grained descriptions lead to analytic solutions of the steady-state
structure \cite{Janssen86}. Ising DDS simulations have been consistent with 
these results \cite{Leung91} on balance, though 
they have been performed mainly at infinite fields \cite{infinitefield}. [There
have been claims that different approaches, though still coarse-grained,
should apply in that infinite-field limit \cite{Marro96,Garrido98}.] 
Away from the critical point, the focus has been on the approach towards the 
final steady state and the question has been whether {\em any} coarse-grained 
model recovers the phenomenology of a stochastic Ising DDS.  In the 
ordered steady-state of both the lattice gas and related continuum
descriptions, the
domain walls align with the field direction and fluctuations are suppressed  to
a remarkable degree \cite{Schmittmann95}.  The main difference occurs in the 
approach to the steady-state, and is easiest to visualize in
a system with a marked minority of one of the ordered phases 
(so-called ``off-critical'' systems). There, separated droplets nucleate
and diffusively coarsens [for the zero-field limit see
\cite{Lifshitz61}].  As the drops grow, they elongate in the field direction ---
ultimately forming the stripes of the steady-state.  This phenomena is seen
both in the lattice DDS \cite{Alexander96} and in continuum models
\cite{Yeung93}. However, the drop shapes are triangular and {\em point in
opposite directions} with respect to the field in the Ising and coarse-grained
models (see, however, \cite{highfield}).  As seen in Fig.\ \ref{FIG:offcrit},
the tips of the triangles point against the field direction (up) in the  lattice
model, while they point with the field (down) 
in the continuum simulations. The same triangular
anisotropy, and discrepancy, is seen at other volume fractions as well.

Does this indicate that either the Ising DDS models, or the 
coarse-grained models (or both!) are non-universal? 
Either would be less than ideal, since stochastic (Ising) models are 
needed for temperature-dependent studies near the critical point, and
deterministic coarse-grained models are both 
analytically tractable and more computationally efficient at lower
temperatures.  An attractive resolution would be that there are regimes of 
parameter space in {\em both} stochastic Ising and deterministic
coarse-grained models which exhibit either sign of triangular anisotropy,
down or up with respect to the
field, with the previously studied models being in the two different regimes.
We take that as a working assumption, and try to flip the anisotropy by
varying model parameters in the Ising and coarse-grained approaches.  We 
report partial success, with asymptotic
reversal in the Ising DDS and reversal up to intermediate times in the 
coarse-grained DDS, and we have 
by no means exhausted the phase-space of model parameters.

Even if we achieve full success, fundamental questions remain. Is the
triangular anisotropy simply a non-universal amplitude, or does it lead to 
different phenomenology?  How do we understand the temperature and field
dependence of the anisotropy, and how does the observed anisotropy in the 
dynamic correlations reflect anisotropies 
due to the surface tension, particle mobility, and
the external field? 
Reconciling the Ising and coarse-grained approaches is simply the
start of the story.  

\subsection*{Lattice Gas Dynamics}

The basic discrete DDS model \cite{Schmittmann95,Leung91,Alexander96}
is an extension of an Ising model 
with conserved Metropolis dynamics:  particles hop with probability
\begin{equation}
	W = Min \left[1, \exp \left( -\beta \Delta H \right) \right],
\end{equation}
where the energy difference $\Delta H= \Delta H_{\rm Ising} + \Delta H_{\rm
Field}$ includes an applied field. Restricting ourselves to a two-dimensional
square lattice, $\Delta H_{\rm Ising}$ is the standard nearest neighbor Ising
Hamiltonian in which a particle interacts with its four nearest sites. [We
absorb $J/k_B$ into the temperature, which we then
measure with respect to the zero-field Ising $T_c$.] 
$\Delta H_{\rm Field}$ is $+E$ if the particle moves one lattice unit 
opposite the field direction
and $-E$ if the particle moves in the field direction, where $E$ is the field
strength.  The dynamics are conserved, so particles hop rather than being
created nor destroyed. Restricting the hops to 
nearest neighboring sites, Alexander et al.\ \cite{Alexander96} 
found that the domains formed upward pointing triangles, as shown in
Fig.~\ref{FIG:offcrit}.

Natural generalizations of this well studied DDS model include looking at 
different lattices [triangular, hexagonal, and Kagom\'{e} in $2d$], 
rotating the field away from a lattice direction \cite{rotate}, 
allowing hops and/or interactions with 
further neighbor particles, and allowing anisotropic hop rates and
interactions.  Universal results on the basic model should be 
robust to these sorts of microscopic differences, as differences will
certainly be entailed by experimental realizations. [These changes 
will affect variously the anisotropic particle
mobility and interfacial surface tension of any coarse-grained
representation.] In this paper we allow next-nearest-neighbor
hops, where we treat all eight immediate neighbors with equal weight. We
label these ``nnn'' dynamics, in contrast to ``nn'' dynamics where hops
are restricted to the four nearest-neighbors.

\subsection*{Coarse-grained Dynamics}

The simplest coarse-grained dynamics is the time-dependent Ginzburg-Landau
(TDGL) model with a field.  The free energy is given as
\begin{eqnarray}
\label{EQN:free}
	F[{\phi}] &=& \int dr \left[ f(\phi(r))+\frac{1}{2} |\nabla \phi|^2 
	+Ez \phi\right],
\end{eqnarray}
where $\phi( {\bf r}, t)$ is the order parameter and the field ${\bf E}$ 
points down toward lower $z$. Within a uniform phase, we use the following
Flory-Huggins type free-energy density:
$$
	f( \phi ) =
	(1+  \phi) \ln (1+\phi) +
	(1-\phi) \ln (1-\phi) - \frac{a}{2} \phi^{2}.
$$
The system will phase separate for $a > 2$ with coexistence values
depending on $a$.  [For $a \approx 2$  this recovers a more familiar
$\phi^{4}$ free energy.] This choice of $f(\phi)$ forces $| \phi | < 1$,
which simplifies the treatment of the particle mobility (below). 

We choose standard TDGL 
dynamics driven by gradients in the chemical potential, so that 
the particle current is 
\begin{eqnarray}
		\vec{J} &=& -M(\phi) \nabla \frac{\delta F}{\delta \phi},
\end{eqnarray}
where $M(\phi)$ is an order parameter dependent mobility.  
A continuity equation is then used to determine the evolution of the 
order parameter:
\begin{eqnarray}
		\frac{\partial \phi}{\partial t} &=& -\nabla \cdot \vec{J} \nn \\
		&=& \nabla
			\cdot M(\phi) \left( \frac{df}{d\phi}-\nabla^2 \phi \right)
			+E \frac{\partial M(\phi)}{\partial z}.
\end{eqnarray}
The  choice of a constant mobility $M(\phi)=M_0$ leads to the field dependence
dropping out of the dynamics.  The next simplest choice,
$M(\phi)=M_0 (1-\phi^2)$, the exact mobility for {\em non-interacting}
lattice-gases, leads to a non-trivial field-dependent DDS coarsening
\cite{Alexander96,Kitahara88,Yeung92}. Indeed,
because the dynamics are deterministic, a semi-quantitative understanding
can then be reached for the linear stability of interfaces and other interfacial
properties \cite{Yeung93}. More generally, we want the mobility to reflect the 
effective coarse-grained mobility. Starting from a
stochastic model with an applied field, the coarse-grained 
mobility will generally be anisotropic, as can be seen explicitly near
the critical point \cite{Janssen86}.  As a minimal step, we allow 
for different mobilities for currents in the $x$ and $z$ directions with
\begin{eqnarray}
\label{EQN:eliptical}
	M_{x}(\phi) & = & (1+m) M_{o} ( 1 - \phi^{2} ),
		\nonumber \\
	M_{z}(\phi) & =  & M_{o} ( 1 - \phi^{2} ), 
\end{eqnarray}
where $m$ describes the mobility enhancement transverse to the 
field direction \cite{onesmallbit}. In the simulations presented here we take
$a \approx 2.75$, so that the bulk phases are at $\phi = \pm 0.8$. We
always set $M_0=1$, which fixes the overall timescale. The initial
conditions $\phi({\bf r},0)$ follows a Gaussian distribution around
$\langle \phi \rangle$.

\subsection*{Asymmetry Measure}

\begin{figure}

\centerline{ \psfig{figure=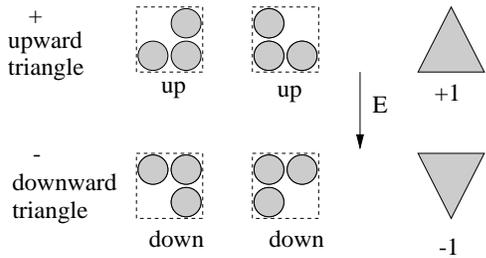,width=2.5in} }

\caption{
The normalized asymmetry
measure is $asym = (n_{up} - n_{down})/(n_{up} + n_{down})$, where
the configurations on top contribute to $n_{up}$ and 
those below contribute to $n_{down}$. }
\label{FIG:asymmeasure}
\end{figure}

In order to quantitatively
compare the models, we need a measure of triangular anisotropy.
We use the microscopic measure 
shown in Fig.~\ref{FIG:asymmeasure}.  That is, we examine all squares
of four nearest-neighbor sites on our lattice and define $n_{up}$
as the number of squares in which
the bottom two sites are positive but the top two sites have opposite
signs.  These configurations point ``upward''.
A similar definition is used for $n_{down}$.  The normalized
asymmetry measure is then
\begin{equation}
	asym = \frac{ n_{up} - n_{down} }{ n_{up} + n_{down} },
\end{equation}
so that $asym = 1$ if all triangles are upward pointing and
$asym = -1$ if all triangles are downward pointing. 
Squares with more or less than three filled sites are not counted. 
The same measure is used for the continuum model except we look
at four neighboring mesh points and count `full' and `empty' as 
$\phi>0$ and $\phi<0$, respectively.
[In practice an equivalent 
asymmetry measure for the `empty' phase can be constructed. Similar results
are obtained.]
Our measure is quantitatively different from that of Alexander {\em et al}
\cite{Alexander96} where normalization drives their asymmetry towards zero 
as domains grow larger, however we obtain the same qualitative sign of the 
triangular asymmetry. We prefer our 
measure since it only depends on the shapes of triangles and not their size,
at least in the coarse-grained formulation. It qualitatively
agrees with anisotropies seen ``by eye''.

\begin{figure}

\psfig{figure=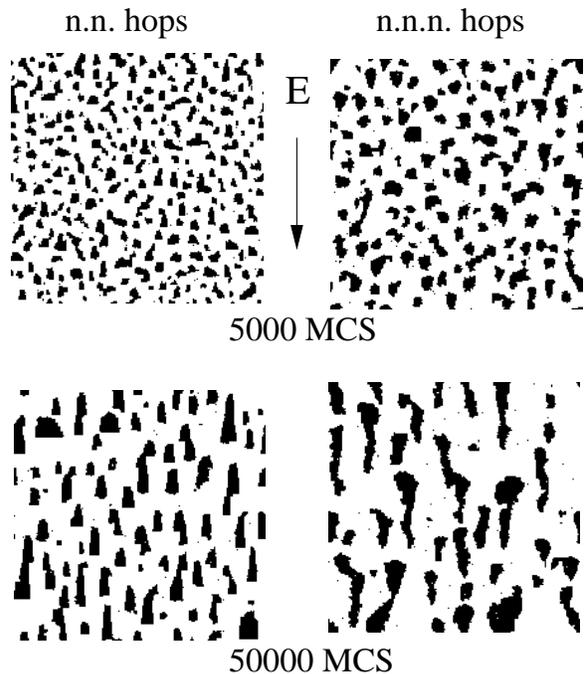,width=3in}

\ \\
\caption{Configuration snapshots of the Ising DDS with
nn (nearest neighbor) and nnn (next nearest neighbor) hops.  The triangles
point in opposite directions with respect to the field. In both cases, 
$\beta E=0.5$ and $T=0.5T_c$. 
\label{FIG:snapIsing}}
\end{figure}

\subsection*{Lattice Gas Results}

\begin{figure}

\centerline{ \psfig{figure=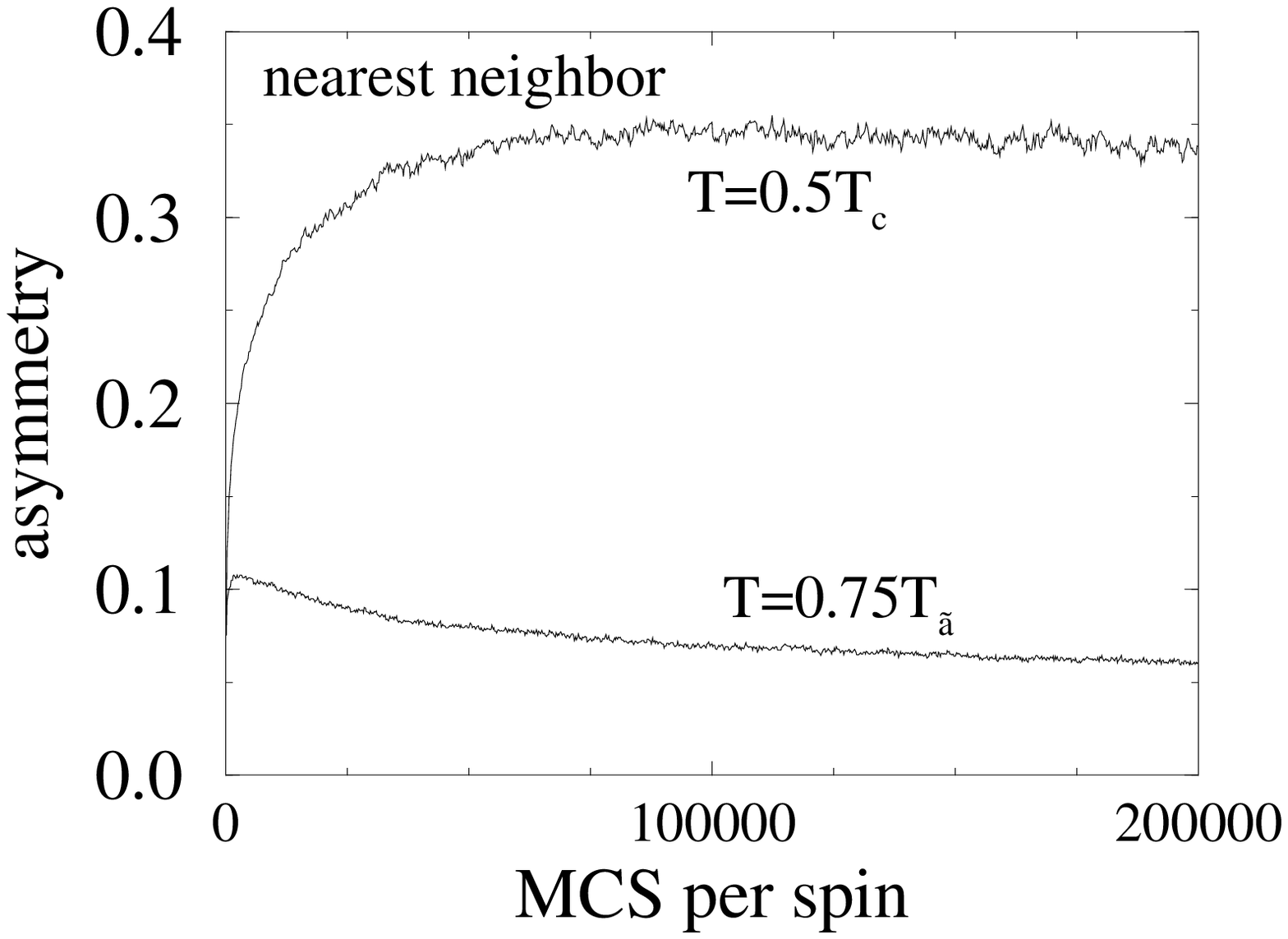,width=2.5in} }

\centerline{ \psfig{figure=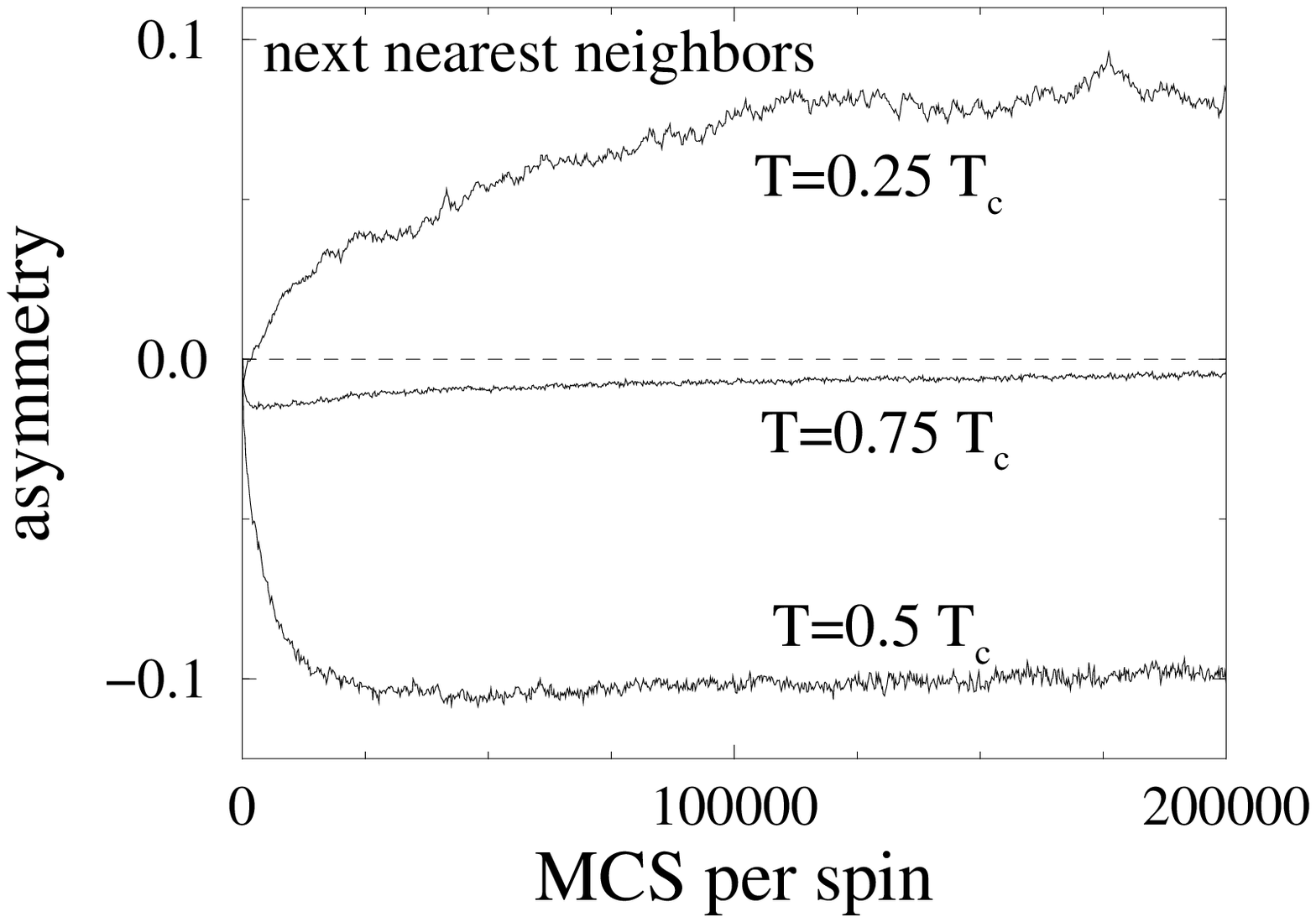,width=2.5in} }

\caption{
The  asymmetry measure as a function of time for the Ising model with
nn and nnn hops, where $\beta E = 0.5$.
The measure seems to approach an  asymptotic value.
\label{FIG:timeIsing}}
\end{figure}

By including next nearest neighbor jumps we can flip the direction of the 
triangular domains with respect to the field. Fig.~\ref{FIG:snapIsing} 
illustrates our results for the Ising model with nn and nnn hops.
This indicates that the sign of the triangular anisotropy is
non-universal. We also observe that the evolution is more rapid when nnn
hops are allowed, as probed by domain size.

The quantitative evolution of the  asymmetry as a function  of time for both nn
and nnn hops is shown in Fig.\ \ref{FIG:timeIsing}.
The data is averaged over 5 to 10 configurations to reduce the noise. 
The asymmetry starts small and eventually
saturates to an asymptotic value. We cannot rule out further change [indeed
slight decay is evident for $T=0.75 T_c$] since there is no known
dynamical scaling in the correlations, i.e. 
no time-independent scaling function. 

\begin{figure}

\centerline{ \psfig{figure=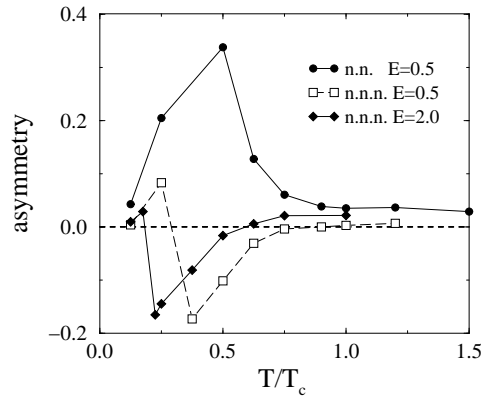,width=2.5in} }

\caption{
The asymptotic  asymmetry vs.\ temperature for the Ising model.
The nn model asymmetry is always positive, while the nnn model 
asymmetry is positive near the critical temperature and at the lowest 
temperatures.  Some field dependence is
also shown. 
\label{FIG:asympT} }
\end{figure}

The asymptotic  asymmetry vs.\ $T/T_c$,
where $T_{c} \equiv T_c(0)$ is the critical temperature for
zero field, is shown in Fig.\ \ref{FIG:asympT}.
With only nn hops the asymmetry
is always positive, indicating upward pointing triangles, and increases
with decreasing temperature.  
We note that the anisotropy measure is {\em continuous} across $T_c(E)$ (which
ranges from $T_c(0)$ to approximately $1.4 T_c(0)$ at 
$E=\infty$ \cite{Leung91}).  
At $T_c(E)$ we would expect surface tensions and particle mobilities to be
isotropic in a small field, and so we attribute the residual anisotropy to
the field. [We note that long-range correlations can persist into the 
high-temperature disordered phase due to violations in detailed balance in 
driven systems \cite{Schmittmann95}.]

With nnn hops, the triangular 
asymmetry is small and  positive near the critical point. 
This is consistent with the previous discussion of nn hops.  At lower
temperatures the asymmetry turns negative. Since the nnn hops should
lead to a more isotropic particle mobility, we might infer that the increasing
{\em anisotropy} of the surface tension with decreasing temperature feeds 
a negative triangular asymmetry when not sufficiently
`compensated' by an anisotropic
mobility.  Clearly the situation is complicated since to 
fully characterize the `anisotropy' requires the entire function of, e.g.,
surface tension vs interfacial orientation. Without actually having
a quantitative measure of the coarse-grained properties of the system
(surface tension, particle mobility) it is difficult to discuss the exact
origins
of the triangular asymmetry. Indeed, {\em this difficulty
is a primary motivation to explore the
coarse-grained picture}. Regardless, the particle mobility will depend on 
the microscopic structure, which in turn depends on the applied field, 
even at $T=0$.  In the limit of small applied field $E$ 
additional induced anisotropies should become negligible. We see some 
indications of this through the reduction in the low-temperature positive 
asymmetry regime with nnn hops as the field is reduced, indicating 
that the regime is induced by the finite field. This suggests one possible
simplifying tactic: to look at the small field limit. Unfortunately this
makes the timescales for numerical investigation of the driven system
inaccessibly large. Analytically, this limit has been profitably
used by one of us in a coarse-grained analysis of surface instabilities
in nearly isotropic systems \cite{Yeung93}.

\subsection*{Coarse-grained Results}

The results for the kinetic Ising model indicates that the positive asymmetry
measure may be due to anisotropy in the mobility.  Therefore to flip the
triangles positive 
in the TDGL model, we allow for different mobilities in the the $x$
and $z$ (field) direction with $M_{x}/M_{z} = 1 + m$.  We also varied the bulk
coexistence value, the field strength $E$ and the
initial filling fraction.  We found that the 
primary effect came from $m$ and from $E$.

\begin{figure}
\centerline{ \psfig{figure=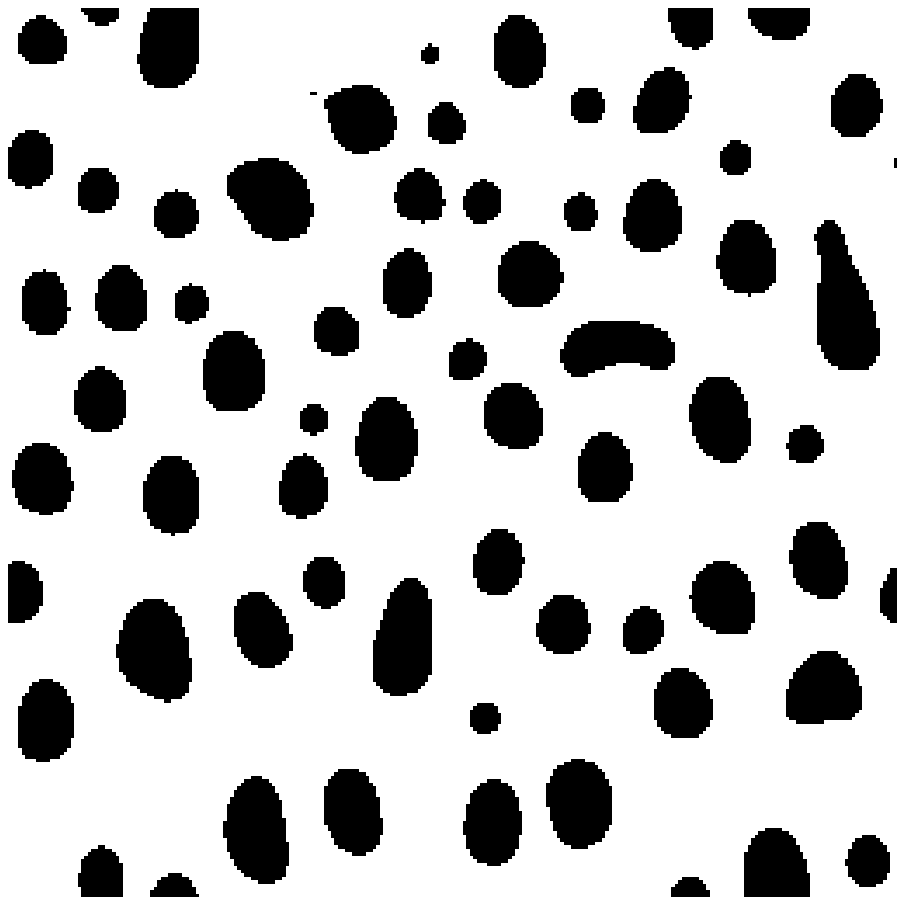,width=2.25in} }
\centerline{ \psfig{figure=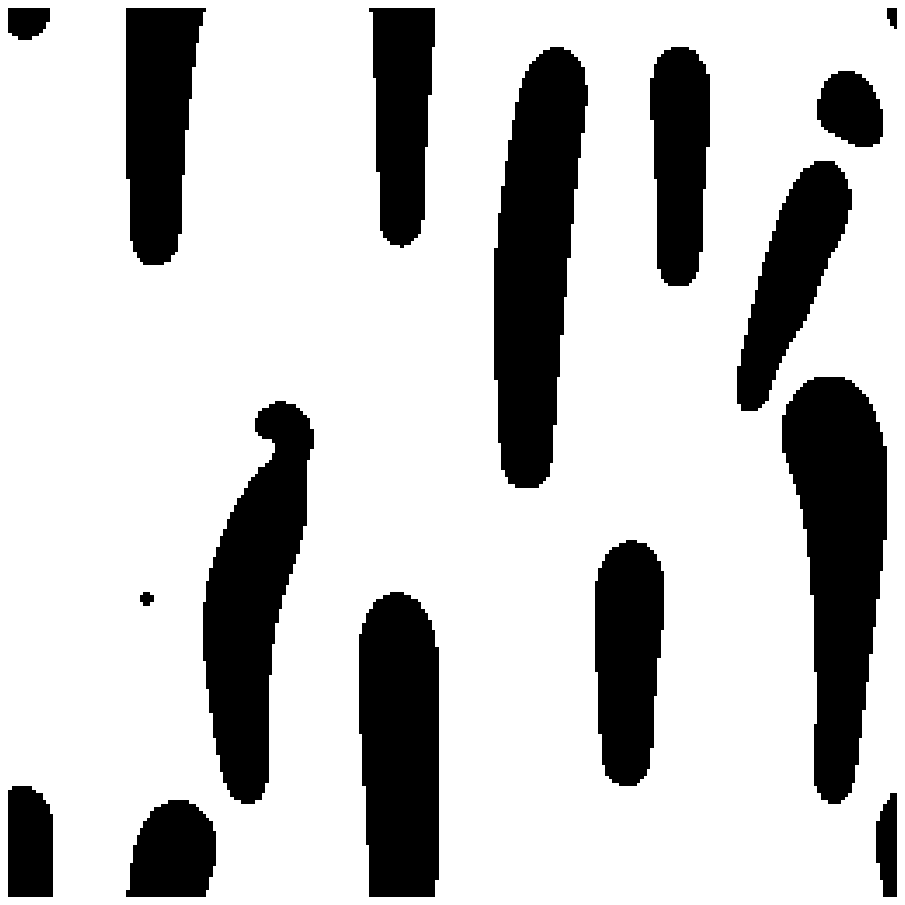,width=2.25in} }
\caption{
Configuration snapshots at $t=1600$ and $t=12800$ 
of an evolving coarse-grained system with $m=1$ and 
$E=0.1$. The reversal of the asymmetry, while not dramatic, can be seen.
The applied field is downward in both snapshots.
\label{FIG:asymswap}}
\end{figure}
\begin{figure}

\centerline{ \psfig{figure=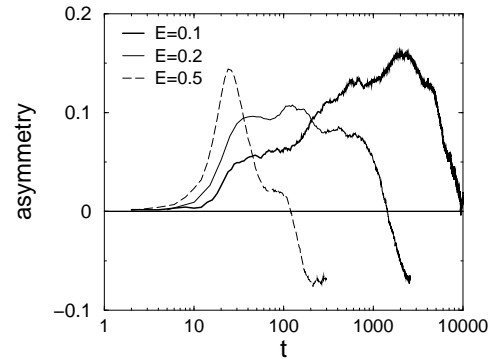,width=2.5in} }

\caption{
The asymmetry as a function of time for fixed mobility anisotropy,
$m = 1$ and different fields.  The asymmetry starts positive
but then becomes negative at late times.  The time for which the
asymmetry is positive is extended for smaller fields.
\label{FIG:asymtcg}
}
\end{figure}

The top snapshot in  Fig.\ \ref{FIG:asymswap} shows that the triangles are in
the same direction as the n.n.\ kinetic Ising model at early times. This is
confirmed by a positive asymmetry measurement at these times. However, the
bottom snapshot in Fig.\ \ref{FIG:asymswap} shows that the asymmetry measure
becomes negative at late times when the domains are very elongated in the field
direction.  This transient 
behaviour was robustly present at all nonzero
values of m and E that we tried.  However, the 
early time ``transient'' regime with
positive asymmetry is quite large and can be extended indefinitely in the limit
of $E/m \to 0$.  This is shown in Fig.\ \ref{FIG:asymtcg} which shows the
asymmetry vs.\ time for critical quenches at fixed $m=1$ and varying $E$. 
This raises the intruiging
question of whether the asymmetries seen in the Ising DDS
models, where the underlying dynamics are slower, might switch at later
times.

\subsection*{Conclusion}

We have shown that the sign and magnitude of triangular anisotropies of
growing domains are non-universal in {\em both} stochastic Ising and
deterministic coarse-grained DDS models. Hence, we see no qualitative
differences between these approaches with finite fields, and rather see great 
promise in using the strengths of each approach to explore DDS
phenomenology.

Questions remain concerning the origins of the triangular anisotropy
within a full anisotropic coarse-grained model. This must be understood
to intelligently explore the parameter space of coarse-grained models. 
With this understanding, we might even profitably turn the tables and
use the triangular anisotropy as a probe of interfacial properties.

We are not overly concerned with the apparent transient nature of the
``flipped'' anisotropy in the coarse-grained model, since even the
stochastic Ising model has not yet been extensively enough studied to tell
if the anisotropies hold asymptotically late. However, we will now focus
our efforts on prolonging the reversal in coarse-grained models. This 
provides a motivation to more fully understand the origins of the triangular
anisotropy. In the process we would like to develop a more intrinsically
coarse-grained measure of anisotropy that can be used equally well in Ising
and coarse-grained approaches. We expect that anisotropies in the Porod tail
of the structure factor will be the most robust measure, 
since they directly probe the distribution of interfacial orientations.

\acknowledgements

A. D. Rutenberg thanks the NSERC, and {\it le Fonds pour la Formation
de Chercheurs et l'Aide \`a la Recherche du Qu\'ebec}. 
C.\ Yeung gratefully acknowledges support for this work from
the Research Corporation under Cottrell College
Science Grant CC3993. We would like to thank Royce Zia for
encouragement and discussions.


\begin{thebibliography}{99}
\bibitem{infinitefield}
The infinite field, $E=\infty$, limit in a lattice DDS is singular since no
currents can point against the field. This constraint on the currents
will then persist in any accurate coarse-grained representation, though
the coarse-grained field must be finite to produce finite currents. 
We might expect some non-universal results from Ising DDS 
in this limit, both in critical phenomena and away from $T_c(\infty)$. 
Indeed, while simulations of critical phenomena  with $E=\infty$ 
\protect\cite{Leung91} are consistent with the coarse-grained
field-theoretic results with a finite field, some debate remains
\protect\cite{Marro96,Garrido98}. Our resolution is simple: 
we restrict ourselves to finite fields.
\bibitem{onesmallbit} We do not cover the entire phase-space
of deterministic coarse-grained models. In general one requires
a fully anisotropic surface-tension, which can be selected by higher-order
gradients in Eqn.~\protect\ref{EQN:free}. One also can require a fully
anisotropic particle mobility, rather the simple elliptical anisotropy of 
Eqn.~\protect\ref{EQN:eliptical}, with more general order-parameter
dependence as well.
\end{thebibliography}
\end{document}